\documentclass{amsart}

\usepackage{arxiv}

\usepackage[utf8]{inputenc} 
\usepackage[T1]{fontenc}    
\usepackage{hyperref}       
\usepackage{url}            
\usepackage{booktabs}       
\usepackage{amsfonts}       
\usepackage{nicefrac}       
\usepackage{microtype}      
\usepackage{lipsum}		
\usepackage{graphicx}
\usepackage{natbib}
\usepackage{doi}
\usepackage{xcolor}
\usepackage{listings}
\definecolor{backcolour}{rgb}{0.90,0.90,0.95}

\usepackage{attrib}
\usepackage{cleveref}

\lstdefinestyle{style}{
backgroundcolor=\color{backcolour},
basicstyle=\ttfamily,
tabsize=2
}
\usepackage{amsthm}

\title{$k$-means considered harmful:\\On arbitrary topological changes in Mapper complexes}
\renewcommand\shorttitle{$k$-means considered harmful}

\usepackage{authblk}
\author[1,2]{Mikael Vejdemo-Johansson \\ \url{mvj@math.csi.cuny.edu}}
\affil[1]{Department of Mathematics, CUNY College of Staten Island}
\affil[2]{Computer Science; Data Science; Mathematics, CUNY Graduate Center}

\hypersetup{
pdftitle={k-means considered harmful},
pdfsubject={TDA},
pdfauthor={Mikael Vejdemo-Johansson},
pdfkeywords={TDA, Mapper, Parameter selection, k-means clustering},
}

\begin{document}
\maketitle

\begin{abstract}
The Mapper construction is one of the most widespread tools from Topological Data Analysis.
There is an unfortunate trend as the construction has gained traction to use clustering methods with properties that end up distorting any analysis results from the construction.
In this paper we will see a few ways in which widespread choices of clustering algorithms have arbitrarily large distortions of the features visible in the final Mapper complex.
\end{abstract}


\clearpage


Mapper was introduced by Gurjeet Singh in 2007 \cite{singh2007topological}, and gave rise to the data analysis company Ayasdi \cite{lum2013extracting}.
Over the years, and especially after Ayasdi pulled back on their academic collaboration programs, open source Mapper implementations have risen: Python Mapper \cite{mullner2013pymapper}, Kepler Mapper \cite{KeplerMapper_JOSS} included in scikit-TDA, Giotto-TDA \cite{tauzin2020giottotda}, and recently tda-mapper \cite{simi_2024_14194667}.

A number of instructions, tutorials and examples in the open source family of Mapper implementations pick either the DBSCAN clustering algorithm or some version of $k$-means -- or other clustering algorithms with fixed target numbers of clusters. In this paper we will explore various ways in which the fixed count choices in particular are a problem.

The ambition of Mapper as described and used is to provide a topological model of the data analyzed that has some degree of fidelity to the shape the data traces out in the ambient space of the raw data.
As observed in \cite{vejdemo2020certified}, breaking the assumptions of the nerve lemma can produce arbitrary changes to the topology of the model over the source data used: homological features can be both created and removed by breaking these assumptions.

Later on, \cite{alvarado2024graphmappergraph} observe that any graph can be realized as a one-parameter mapper graph, and indeed any small enough simplicial complex can be realized as a mapper complex (with vertices $V$, simplices $K$ and isolated vertices $I$) over any large enough dataset $X$: as long as $|X| \geq |K\setminus V| + |I|$.

Since $k$-means with a globally set $k$ nearly certainly will be producing bad covers (in the sense of the nerve lemma), these choices of clustering algorithms are likely to produce misleading results and should be avoided.


\section{The Mapper Algorithm}

Recall the Mapper algorithm:

\paragraph*{Inputs:} A dataset $X\subset A$, a function $L:X\to Y$, a covering $Y = \bigcup_i Y_i$, a clustering method $\hat\pi_0:2^A\to \hom_{\textrm{Set}}(X, \mathbb{N})$ that sends a set $X$ in $A$ to a method for labeling points in $X$ with cluster id numbers.

\begin{enumerate}
    \item Pull back the covering by $Y_i$ to form a covering $\Xi_i = L^{-1}Y_i$ of $X$.
    \item Refine the covering $\Xi_i$ by introducing $X_{i,j}=(\hat\pi_0(X_i))^{-1}(j)$ -- or in words, let each cluster within a fiber of a cover element be a separate set in a refined covering.
    \item Create the Mapper complex as the Nerve complex $\mathcal{N}(X_{\ast,\ast})$ of all these refined cluster elements. 
\end{enumerate}

In this paper, we will pay particular attention to the choices of $\hat\pi_0$, and the ways in which these choices can reduce the usefulness of the resulting Mapper complex.

\section{The $k$-Means Algorithm}

$k$-means is a clustering algorithm that proposes $k$ cluster means, and iteratively improves the choice of cluster means, optimizing for placing these points so that their Voronoi cells split the data into the chosen number of clusters. The optimization penalizes high distances between points within a cluster and low distances between points that belong to different clusters.

Crucially, $k$-means is a fixed count clustering algorithm: it will produce $k$ clusters for the user. Whether or not the data is suitable to describe as $k$ separated clusters is not considered, or detected, by the algorithm in question.

\section{Uses of $k$-Means}

We have found $k$-means (and other fixed count methods, such as agglomerative clustering in \texttt{scikit-learn}) in tutorials, documentation and example code in the Open Source Mapper community:

\begin{quote}
For clustering we use Agglomerative Single Linkage Clustering with the “cosine”-distance and 3 clusters.
\attrib{Kepler-Mapper and NLP examples}
\end{quote}

\begin{quote}
    \begin{verbatim}
# Define the simplicial complex
scomplex = mapper.map(lens, X, cover=km.Cover(n_cubes=15, perc_overlap=0.7),
clusterer=sklearn.cluster.KMeans(n_clusters=2, random_state=3471))        
    \end{verbatim}
    \attrib{Kepler-Mapper demo "Choosing a lens"}
\end{quote}

\begin{quote}
    \begin{verbatim}
# Use KMeans with 2 clusters
graph = mapper.map(X_projected, X_inverse, 
clusterer=sklearn.cluster.KMeans(2))        
    \end{verbatim}
    \attrib{Kepler-Mapper documentation for the main API command}
\end{quote}

\begin{quote}
    \begin{verbatim}
mapper_algo = MapperAlgorithm(
  cover=CubicalCover(
    n_intervals=10, overlap_frac=0.65 
  ), 
  clustering=AgglomerativeClustering(10),
  verbose=False )        
    \end{verbatim}
    \attrib{\texttt{tda-mapper} documentation, digits dataset example}
\end{quote}

\section{The critique}

The warnings in this paper build on a series of preceding and increasingly concerning results pointing to weaknesses in the way our community uses the Mapper algorithm.

There are two core ways in which we can draw conclusions from a Mapper analysis:
\begin{enumerate}
    \item \textbf{Discrimination:} Because it is an invariant, if two datasets give dramatically different results with similar choices of parameters, we may conclude that the datasets are different.
    \item \textbf{Description:} Because of the Nerve Lemma, we may believe that the shape of the resulting Mapper complex is related to the abstract notion of a shape of the data source.
\end{enumerate}

The \textbf{Discrimination} property is a result of stability, and out of scope for this paper; we see no immediate reason to doubt the discriminative ability of Mapper analyses.

For the \textbf{Description} property, an early critique was offered in \cite{vejdemo2020certified}, which observed that failure to produce a \emph{good cover} (in the sense of the appropriate Nerve lemma) can produce arbitrary changes in the topology of the output.

To this point, in 2024, \cite{alvarado2024graphmappergraph} show that any graph, or indeed any small enough simplicial complex can be realized as a mapper complex over any large enough dataset by choosing lens function carefully.

We continue this line of critiques by pointing out that a fixed-count clustering algorithm is all but guaranteed to over-produce clusters in some cover elements and under-produce clusters in others. Over-production means splitting clusters that should stay together and producing holes (as the split parts get connected in the adjacent cover elements) or other topological features, and under-production means forming quotient spaces with arbitrary identification of topological structures in the data.

\begin{figure}
\includegraphics[width=0.2\textwidth]{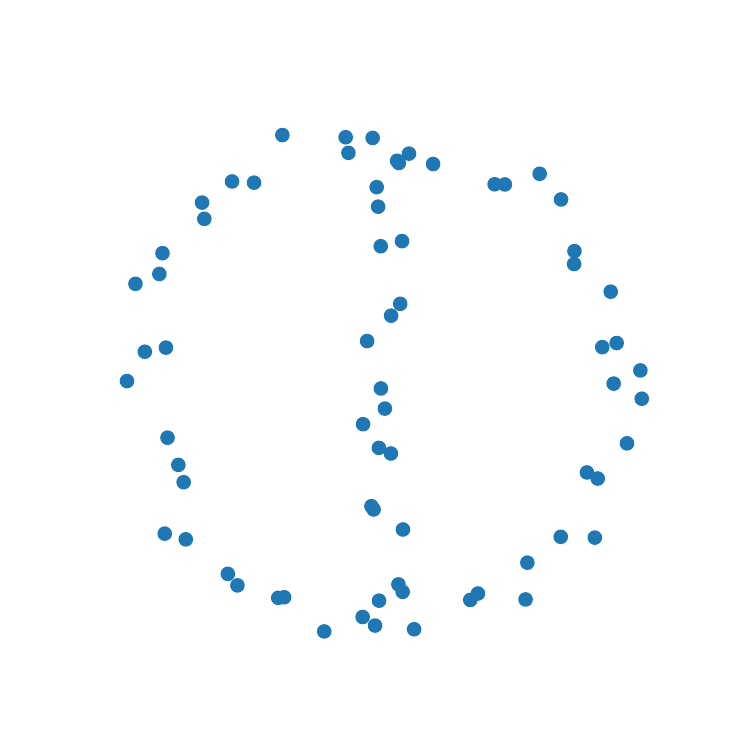}
\includegraphics[width=0.2\textwidth]{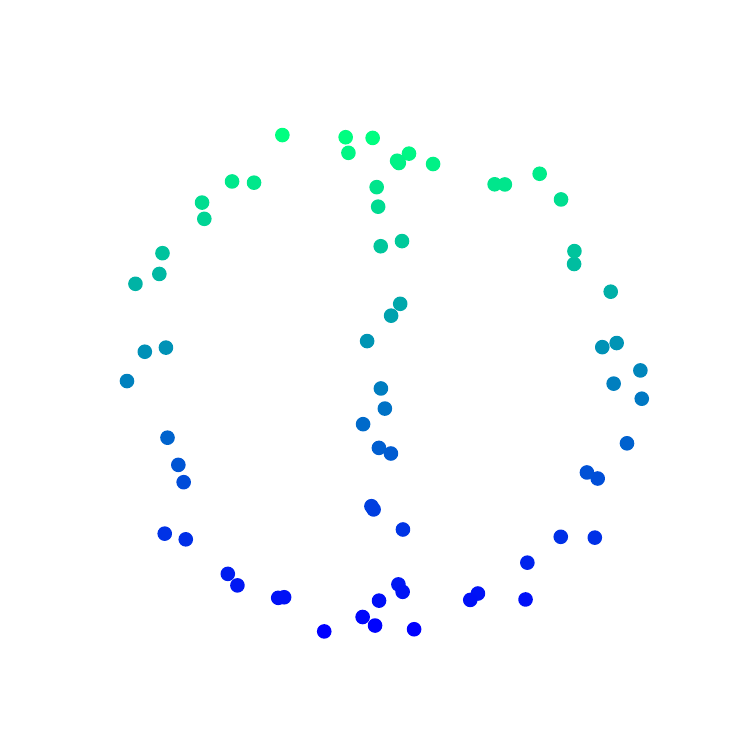}
\includegraphics[width=0.2\textwidth]{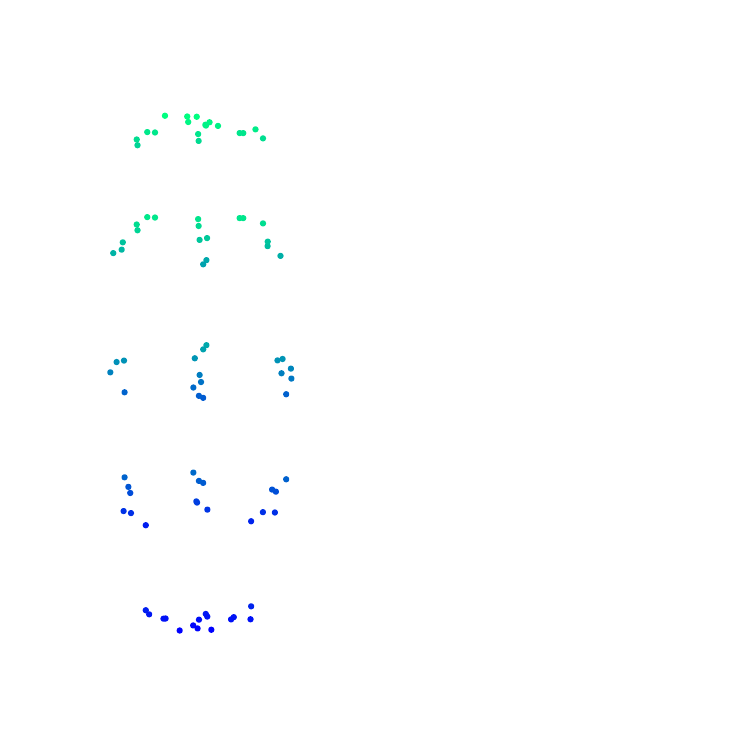}
\includegraphics[width=0.2\textwidth]{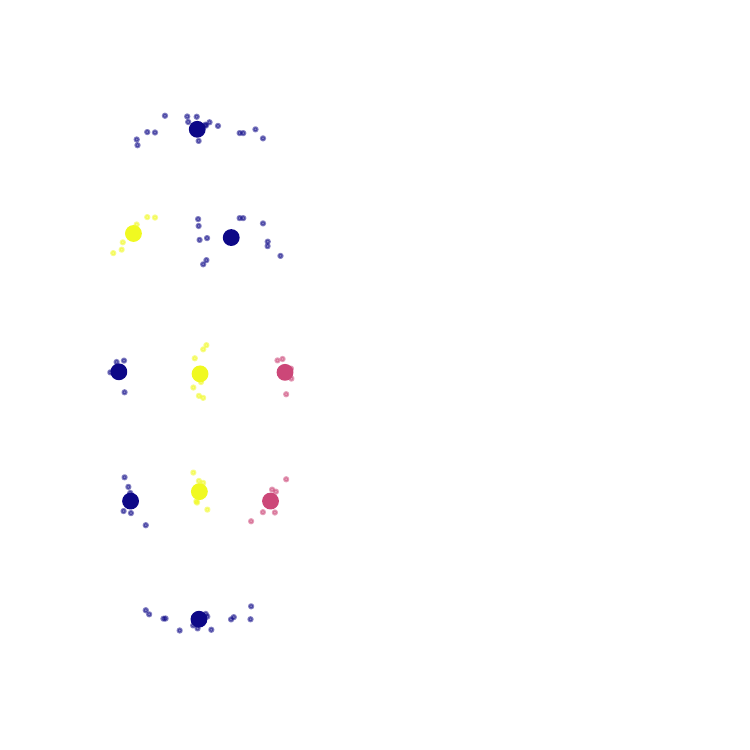}
\\
\includegraphics[width=0.3\textwidth]{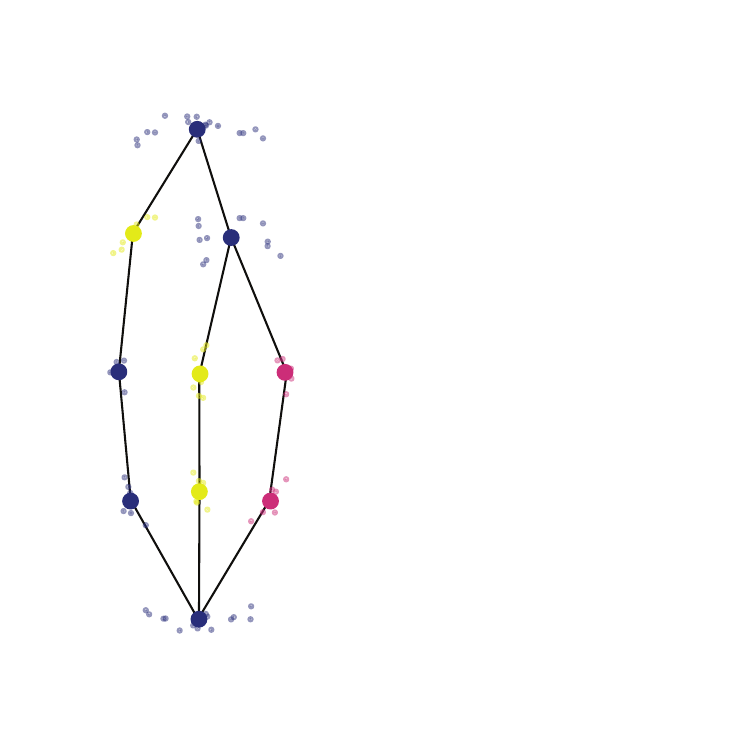}
\,
\includegraphics[width=0.3\textwidth]{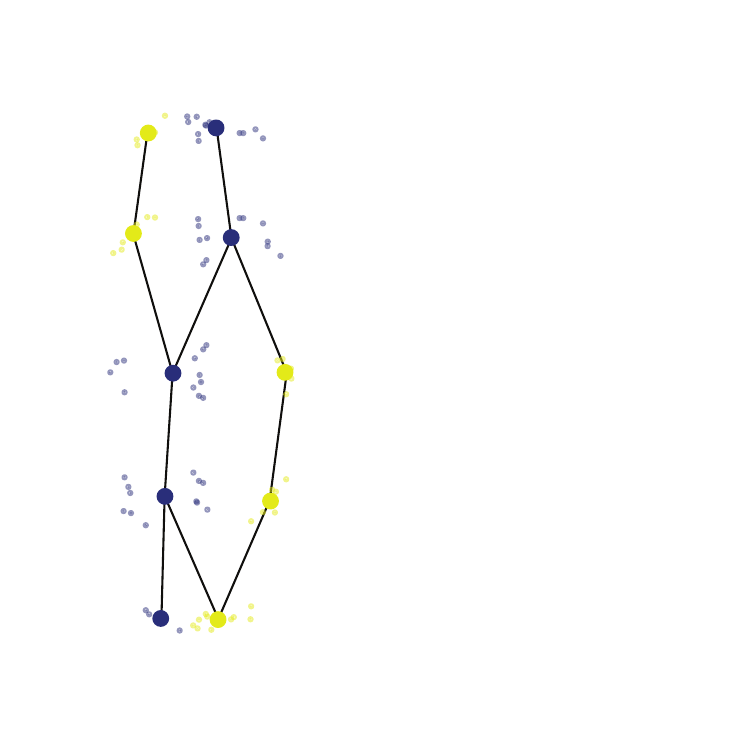}
\,
\includegraphics[width=0.3\textwidth]{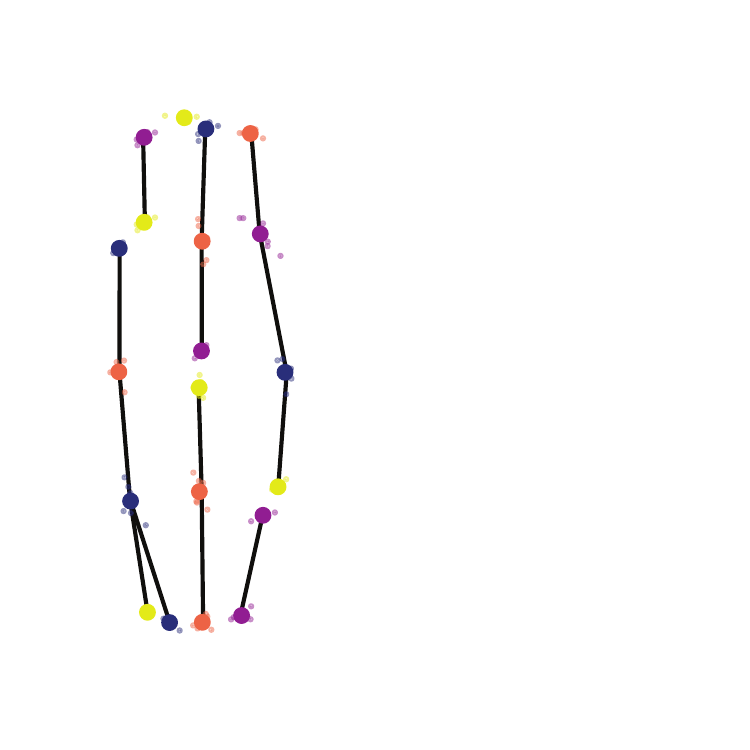}
    \caption{Illustration of the Mapper workflow, as well as illustrations of failure modes from both over-producing and under-producing clusters. In the bottom row the resulting Mapper complexes, from left to right: expected output, output with 2-means, output with 4-means.}
    \label{fig:mapper-example}
\end{figure}

In \cref{fig:mapper-example}, we illustrate both the Mapper algorithm and these failure modes on a concrete example data set. Notice how for the top and bottom covering elements -- where the figure connects fully -- both 2-means and 4-means over-produce clusters, while 2-means under-produces clusters on the intermediate stages, collapsing two parallel strands into a single one.

We do want to point out, and emphasize, that the problems we are pointing out here -- with potentially arbitrarily large changes to the topology of the Mapper complex -- is a product of the mandated choice of the number of clusters when using $k$-means, and how this global choice is guaranteed to either merge or split clusters in the preimages.

The important recognition here is that it is this merging and splitting that forms the problem -- and not merely a problem of uneven point distribution. Hence, a solution based on pre-processing, or on picking a different fixed count clustering solution is not going to solve this problem. On the other hand, an adaptive method that \emph{uses} $k$-means could very well be found to work. If we had an oracle that specified a likely number of clusters for each preimage, and then used $k$-means, quality would improve dramatically, and the failure modes illustrated in our examples in this paper would no longer apply.

This observation favors the largest gap heuristic introduced in the first paper, as well as methods that use clustering quality scores to pick appropriate hyperparameters for each instance.

\section{Possible solutions}

At the core of this critique is the fundamental unreliability of using fixed-count clustering techniques: $k$-means, fixed count agglomerative clustering, and anything else that doesn't adapt the clustering result to the data on hand.
It seems to the author that these choices are often made out of convenience: picking up \texttt{scikit-learn} and picking the simplest algorithm one can find.
One of the most reasonable solutions has been around since the first publications of the Mapper algorithm: Gurjeet Singh and Ayasdi have been using hierarchical clustering with a largest gap heuristic, first described in the original Mapper paper itself \citep{singh2007topological}.
This paper states as required features for a clustering algorithm to be used with Mapper:
\begin{quote}
    Do not require specifying the number of clusters beforehand.
\end{quote}
and describes their approach to clustering as:
\begin{quote}
    If we look at
the histogram of edge lengths in $C$, it is observed experimentally, that shorter edges which connect points within each
cluster have a relatively smooth distribution and the edges
which are required to merge the clusters are disjoint from
this in the histogram. If we determine the histogram of $C$
using $k$ intervals, then we expect to find a set of empty interval(s) after which the edges which are required to merge
the clusters appear.
\end{quote}

This approach was later refined by Gurjeet Singh and Ayasdi Inc., who patented their developed version.
There are surely many different approaches that can work in this setting -- one of the reviewers graciously suggested using the \emph{silhouette score}~\citep{rousseeuw1987silhouettes} as one feasible approach. DBSCAN~\citep{el2004efficient} is another widely spread approach (that requires finesse in handling the category of un-clustered points inherent in the DBSCAN algorithm).

\clearpage

\bibliographystyle{unsrtnat}
\bibliography{references}  

\end{document}